\documentclass[prl,aps,twocolumn]{revtex4-1}
\pdfoutput=1
\usepackage{epsfig}
\usepackage{amsfonts}
\usepackage{rotating}
\usepackage{sparticles}
\usepackage{dcolumn}
\usepackage{bm}
\usepackage{textcomp}
\usepackage{hyperref}
\hypersetup{
    bookmarks=true,         
    unicode=false,          
    pdftoolbar=true,        
    pdfmenubar=true,        
    pdffitwindow=false,     
    pdfstartview={FitH},    
    pdftitle={My title},    
    pdfauthor={Author},     
    pdfsubject={Subject},   
    pdfcreator={Creator},   
    pdfproducer={Producer}, 
    pdfkeywords={keyword1} {key2} {key3}, 
    pdfnewwindow=true,      
    colorlinks=true,       
    linkcolor=red,          
    citecolor=red,        
    filecolor=magenta,      
    urlcolor=green,           
    linktocpage=true
}
\usepackage{amsmath,amssymb,amsthm,graphicx,latexsym}
\usepackage{subfigure}
\usepackage{pstricks}
\usepackage{color}

\usepackage{mathrsfs}
\usepackage{hyperref}

\newcommand{\be}{\begin{equation}}
\newcommand{\ee}{\end{equation}}
\newcommand{\bea}{\begin{eqnarray}}
\newcommand{\eea}{\end{eqnarray}}
\newcommand{\bml}{\begin{subequations}}
\newcommand{\eml}{\end{subequations}}
\newcommand{\bfig}{\begin{figure}}
\newcommand{\efig}{\end{figure}}

\newcommand{\del}{\delta}

\newcommand{\thg}{\theta}

\newcommand{\slashed}{\hspace{-1.1ex}/}

\begin{document}
\title{Quantum randomness in the Sky
}

\author{{{Sayantan Choudhury}}$^{1}$\footnote{Electronic address: {sayantan.choudhury@aei.mpg.de, sayanphysicsisi@gmail.com}} ${}^{}$
	and {{Arkaprava Mukherjee}}$^{2}$
	\footnote{Electronic address: {am16ms058@iiserkol.ac.in
	}} ${}^{}$}
\affiliation{$^1$Quantum Gravity and Unified Theory and Theoretical Cosmology Group, Max Planck Institute for Gravitational Physics (Albert Einstein Institute),
	Am M$\ddot{u}$hlenberg 1,
	14476 Potsdam-Golm, Germany
}
\affiliation{$^2$Department of Physical Sciences, Indian Institute of Science Education and Research Kolkata,
	Mohanpur, West Bengal 741246, India
}

\begin{abstract}
In this article, we study quantum randomness of stochastic cosmological particle production scenario using quantum corrected higher order Fokker Planck equation. Using the one to one correspondence between particle production in presence of scatters and electron transport in conduction wire with impurities we compute the quantum corrections of Fokker Planck Equation at different orders. Finally, we estimate Gaussian and non-Gaussian statistical moments to verify our result derived to explain stochastic particle production probability distribution profile.

\end{abstract}
\maketitle

It is a well known fact that the particle production scenario in the early universe cosmology (during reheating) follows the dynamical master equation, {\it aka} Klein-Gordon equation. On the other hand, transport phenomena of electron through a conduction wire with impurities follow time independent Schr{\"o}dinger equation. Both of this dynamical time dependent phenomena have structural one to one correspondence \cite{Amin:2015ftc,Amin:2017wvc}.  Anderson Localization and saturation of the chaos are some well studied phenomena in the context of scattering problem can be extended to describe the quantum randomness phenomena during cosmological particle production. From their inherent stochastic nature quantum chaos can be related to them and chaos bound can be defined either by Lyapunov exponent \cite{Maldacena:2015waa} or by Spectral Form Factor \cite{Choudhury:2018lcb,Choudhury:2018rjl}. The possible quantum effects arising from higher order corrections in dynamical master equation {\it aka} Fokker Planck equation for particle production scenario in the early universe cosmology (during reheating) can be achieved from the present discussion. For comparing scattering event with stochastic particle production Dirac Delta profile of time dependent coupling (mass function) is chosen, 
\be m^{2}(\tau)=\sum_{j=1}^{N} m_{j}\del_{D}(\tau-\tau_{j}),\ee
localized at time scale $\tau=\tau_j$ (where $j$ represents the number of non-adiabatic events). Further using the concept of transfer matrices occupation number can be computed from this set up. To model a phenomenological situation where width ($w_j$) of the profile of the time dependent coupling is finite and the scattering event is relevant, we consider  {\it sech}  scatterers. It is important to note that, in the limit $w_j\to \infty$ the Dirac Delta profile can be recovered from this phenomenological profile.

In the context of disippative system, {\it Fokker Planck equation} explains the probability density for particle position of Brownian motion in a random system. For a Markovian process this situation can be expressed by {\it Chapman-Kolmogorov equation} \cite{Amin:2015ftc}. Now considering {\it Maximum Entropy Anstaz} we can derive the {\it Fokker Planck equation} from {\it Smoluchowski equation} when we integrate the probability density over the angular coordinate $\theta$:
\be
P(n,\theta,\phi;\tau+\del\tau)\equiv P(n,\theta;\tau+\del\tau)\rightarrow
\langle P(n+\del n ;\tau)\rangle_{\del\tau}~~~
\ee
where we consider an infinitesimal change ($\delta\theta$) is not functionally dependent on $\theta$. Further Taylor expansion of $\langle P(n+\del n ;\tau)\rangle_{\del\tau}$ with respect to the infinitesimal occupation number ($\delta n$) with the constraint in this context $\langle P(n;\tau)\rangle_{\del\tau}=P(n,\tau)$ gives the following result:
\be
  \langle P(n+\del n ;\tau)\rangle_{\del\tau}=\langle P(n;\tau)\rangle_{\del\tau}+\sum^{\infty}_{q=1}(q!)^{-1}\partial^q_{n} P(n;\tau)\rangle_{\del\tau} ~~~~
  \ee
This gives the following general structure of {\it Fokker-Planck equation} which we will use for our all calculations:
\be \label{ez1}
\partial_{\tau} P(n;\tau)=\sum^{\infty}_{q=1}(q!)^{-1}\left(\langle (\del n)^q\rangle_{\del\tau}/\del\tau\right)\partial^{q}_{n}P(n;\tau)~~~~\ee
Using {\it Smoluchowski equation}  the occupation number can be expressed as:
\be
\del n \equiv n_{2}(1+2n)-2 \sqrt{(1+n_{2})(1+n)n_{2}n}~\cos{2(\phi _{2}-\thg)}
\ee
 which help us to further define various statistical moments from the probability density function. Assuming that the particle production rate is small locally ($\mu\delta\tau<1$) we have the truncation in Taylor expansion. With primary truncation in first order $\langle (\del n)\rangle_{\del\tau}$ {\it Fokker-Planck equation} is derived as:
\be \label{ez2}
\mu^{-1}_{k}\partial_{\tau}P(n;\tau)=\partial_{n}\left[ n(1+n) \partial_{n}P(n;\tau) \right]
\ee
Here the mean particle production rate have Fourier mode dependence ($\mu_k$). By Fourier transformation with respect to the occupation number $n$ of the distribution function:
\bea\label{ftv1} P(n;\tau)=(2\pi)^{-1}\int dk~e^{ikn}\bar{P}(k;\tau).\eea
Which simplifies the {\it Fokker Planck equation} in Fourier space :
\bea\label{ftv2} \partial_{\tau} \bar{P}(k;\tau)=\mu_k\left(2ink-k^2n^2\right)\bar{P}(k;\tau),\eea
Imposing initial condition for probability distribution function at time $\tau$ is given by the Dirac Delta profile or its derivatives in different orders we get:
\be \label{boundary1}
\partial^J_{\tau}  P(n;\tau)=(-1)^J ~n^{-J}J!~\delta(n)~~~\forall J=0,1,2,\cdots
\ee
where $J$ denoting the order of quantum corrected {\it Fokker Planck Equation}. 



For $J=1$ we get the following solution of the probability density function
 :
\bea 
	 P(n;\tau)=\frac{\exp[-n(\mu_k(n+1)\tau +\frac{1}{4\mu_k\tau(n +1)}~+1)]}{2\sqrt{\mu_kn(n+1)\tau\pi }}~~~.
	\eea
Comparing the coefficient of $\del\tau $ from the both sides of the Taylor expansion we get quantum corrected {\it Fokker Planck equation} at different order. Without truncation on both sides of this expression additional contributions in $\delta\tau$ and in its higher order can be obtained and generate quantum corrected version of the {\it Fokker Planck equation} valid upto higher orders. All such higher order corrections justify non-Gaussian effects appearing during cosmological stochastic particle production in reheating phase. In another words origin of higher order contributions describe the quantum effects from its non vanishing statistical moments originating from quantum correlations.

Equating both sides of Eq~(\ref{ez1}) after Taylor expansion and comparing the coefficient of $\del\tau ^{2}$ the {\bf \it second order Fokker Planck quation} is computed as:
	\bea\label{ez24} 
&&\left[n^2(1 + n)^{2}/2~\partial^{4}_{n}+2n (1 + 3 n + 2 n^2)\partial^{3}_n\right.\nonumber\\
&& \left.+(1 + 6 n + 6 n^2) \partial ^{2}\right]P(n;\tau) =\mu^{-2}_k \partial^{2}_{\tau} P(n;\tau)
\eea
At at the second order the probability distribution function has the form:
\begin{widetext}
\bea
	 P(n;\tau)&=&\left(\pi  (n^2- \mu ^2_k \tau ^2)\right)^{-1}\left[n \sin (L  n) \cos (L  \mu_k  \tau )-\mu_k  \tau  \cos (L  n)
	 	\sin (L  \mu_k  \tau )\right] -(4 \pi  \mu_k  n)^{-1}\left[i \left\{\text{Ci}(-L  (n+\mu_k  \tau )) \right.\right.\nonumber\\&& \left.\left.
 	\displaystyle
 	~~-\text{Ci}(L  (n+\mu_k  \tau ))\right\} -\text{Ci}(-L  (n-\mu_k  \tau ))+\text{Ci}(L  (n-\mu_k  \tau )) -2 i \left\{\text{Si}(L  (n+\mu_k  \tau ))- \text{Si}(L  (n-\mu_k  \tau ))\right\}\right],~~~~~
\eea
\end{widetext}
where $L$ is the momentum cut-off.

Following the same procedure from Eq~(\ref{ez1}) and comparing the coefficient of $\del\tau ^{3}$ the third order {\it Fokker Planck equation} is obtained as:
\begin{widetext}
	\bea 
&&\left[ n^3(1 + n)^{3}/6~\partial^{6}_n
+ 3n^2(1 + n)^2 (1 + 2 n)/2~\partial^{5}_n
+ 3 n (1 + n) (1 + 5 n + 5 n^2)\partial^{4}_n
\right.\nonumber\\&& \left.~~~~~~~~~~~~~~~+(1 + 2 n) (1 + 10 n + 10 n^2)\partial^3_n\right]P(n;\tau)=\mu^{-3}_k\partial^3_{\tau}P(n;\tau).~~~~~~
\eea
\end{widetext}
Three fold boundary conditions for this equation for J=1,2 and 3 from Eq.~(\ref{boundary1}) with the same initial conditions
we get the following probability distribution function from third order contribution as given by:
\begin{widetext}
	\bea 
		&&P(n;\tau)=\frac{((\sqrt{3}+3 i) \mu_k +2 (\sqrt{3}+i)) n^3}{4 (\sqrt{3}+2 i) \mu ^2_k n^2 ((-1)^{2/3} \mu_k  \tau +n) \sqrt{((-1)^{2/3} \mu_k  \tau +n)^2}}\nonumber\\
		&&+2 i n^2 (2 i \sqrt{3} \mu ^2_k \tau +\mu_k  (\sqrt{-\sqrt[3]{-1} \mu ^2_k \tau ^2+n^2+2 (-1)^{2/3} \mu_k  n \tau }+3 i \sqrt{3} \tau +3 \tau )-2 \sqrt{-\sqrt[3]{-1} \mu ^2_k \tau ^2+n^2+2 (-1)^{2/3} \mu_k  n \tau })\nonumber\\
		&&-\mu_k  n \tau  ((-(\sqrt{3}-3 i)) \mu ^2_k \tau +2 \sqrt[6]{-1} \mu  (\sqrt{-\sqrt[3]{-1} \mu ^2_k \tau ^2+n^2+2 (-1)^{2/3} \mu_k  n \tau }+3 i \sqrt{3} \tau +3 \tau)\nonumber\\
		&&-2 (\sqrt{3}-i) \sqrt{-\sqrt[3]{-1} \mu ^2_k \tau ^2+n^2+2 (-1)^{2/3} \mu_k  n \tau })+(\sqrt{3}+i) \mu ^2_k \tau ^2 (2 \mu_k  \tau +\sqrt{-2 i (\sqrt{3}-i) \mu ^2_k \tau ^2+4 n^2+4 i (\sqrt{3}+i) \mu_k  n \tau }).~~~~~~~
\eea
\end{widetext}
For fourth order contribution equating both sides of Eq~(\ref{ez1}) and comparing the coefficient of $\del\tau ^{4}$ we get fourth order {\it Fokker Planck equation} as given by:
\begin{widetext}
\bea\label{ss12}
&&\left[70 n^4 (1+n)^4\partial^{8}_n+140 n^3 (1+2 n)\partial^{7}_n+30 n^2 (1+n)^2 (3+14 n+14 n^2)\partial^{6}_n \right.\nonumber\\&&\left.
\displaystyle
~~~~+20 n (1+n) (1+2 n) (1+7 n+7 n^2)\partial ^{5}_n+(1+20 n+90 n^2+140 n^3+70 n^4)\partial^{4}_n\right]P(n;\tau)=\mu^{-4}_k\partial^4_{\tau}P(n;\tau).~~~~
\eea
\end{widetext}
Applying four fold boundary conditions (J=1,2,3,4) from Eq.~(\ref{boundary1})
we get the following expression for the probability distribution function, as 
given by:
\begin{widetext}
\bea
	 &&P(n;\tau)=-(2\pi)^{-1}\int^{q}_{-p}dk~e^{ikn}~\left\{\frac{(k^2 n^2 \mu _k^2+2 k n \mu _k+6)}{4 k^3 n^3 \mu _k^3}~ e^{-\mu_k k \tau}+\frac{(k^2 n^2 \mu _k^2-2 k n \mu _k+6)}{4 k^3 n^3 \mu _k^3}~ e^{\mu_k k\tau}\right.\nonumber\\&&\left.
	\displaystyle
~~~~~~~~~~~~~~~~~~~~~~~~~~~~~~~~~~~~~~~~~~~~~~~~~~~~~~~~~~~~~~~~~~~~+\frac{(k^2 n^2 \mu _k^2-6)}{2 k^3 n^3 \mu _k^3} \sin (\mu_k k \tau)+\frac{1}{k^2 n^2 \mu _k^2} \cos(\mu_k k \tau)\right\}
\eea
\end{widetext}
where we introduce IR and UV regulators, $p<k<q$.
\begin{figure}[h!] 
		\includegraphics[width=9cm,height=10cm] {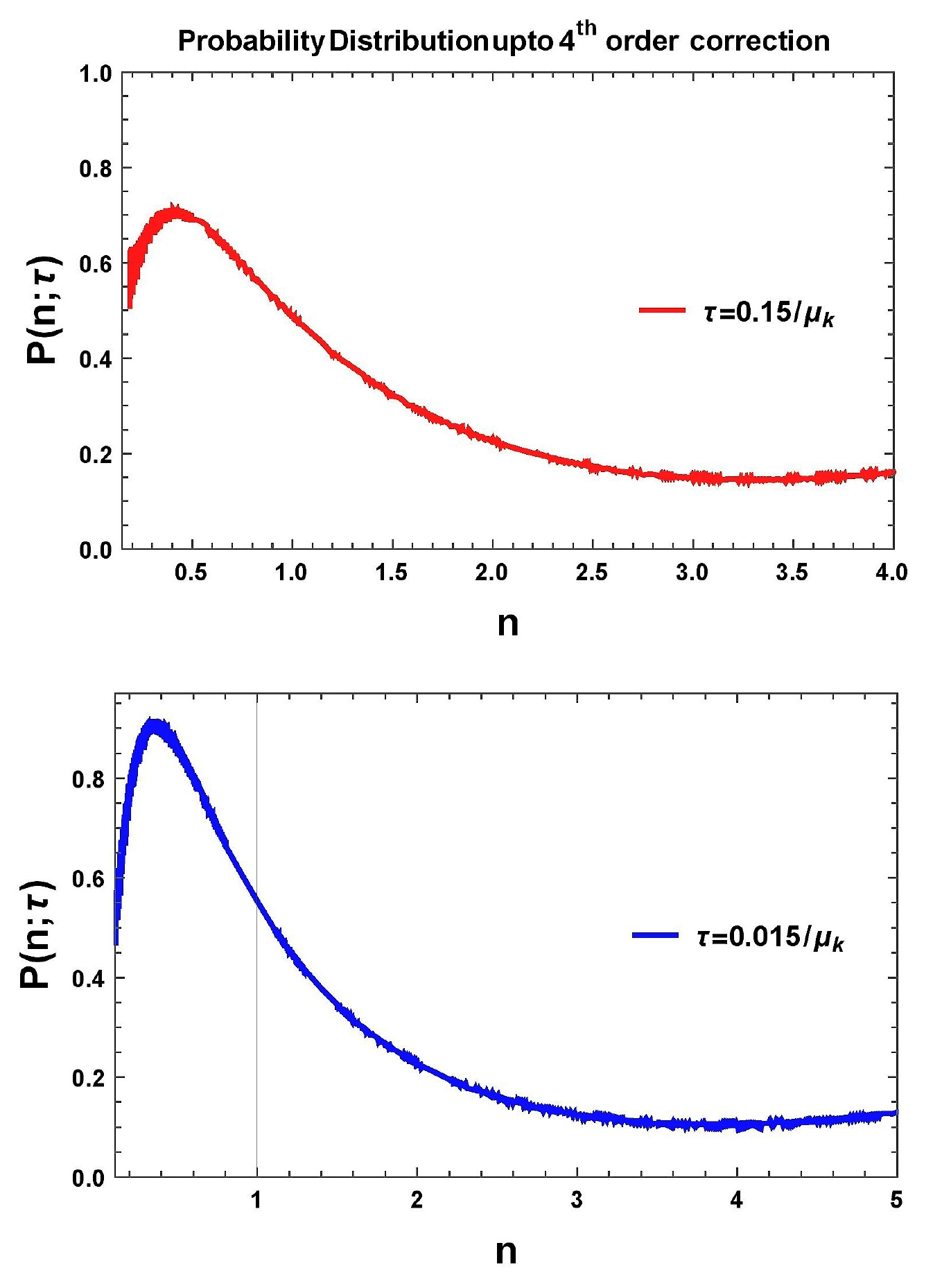}
	\caption{Variation of fourth order quantum corrected probability distribution with respect to the particle number density.}
	\label{Allcorrection}
\end{figure}
\begin{figure}[ht]
\centering
\subfigure[Time evolution of variance. ]{
    \includegraphics[width=8cm,height=5cm] {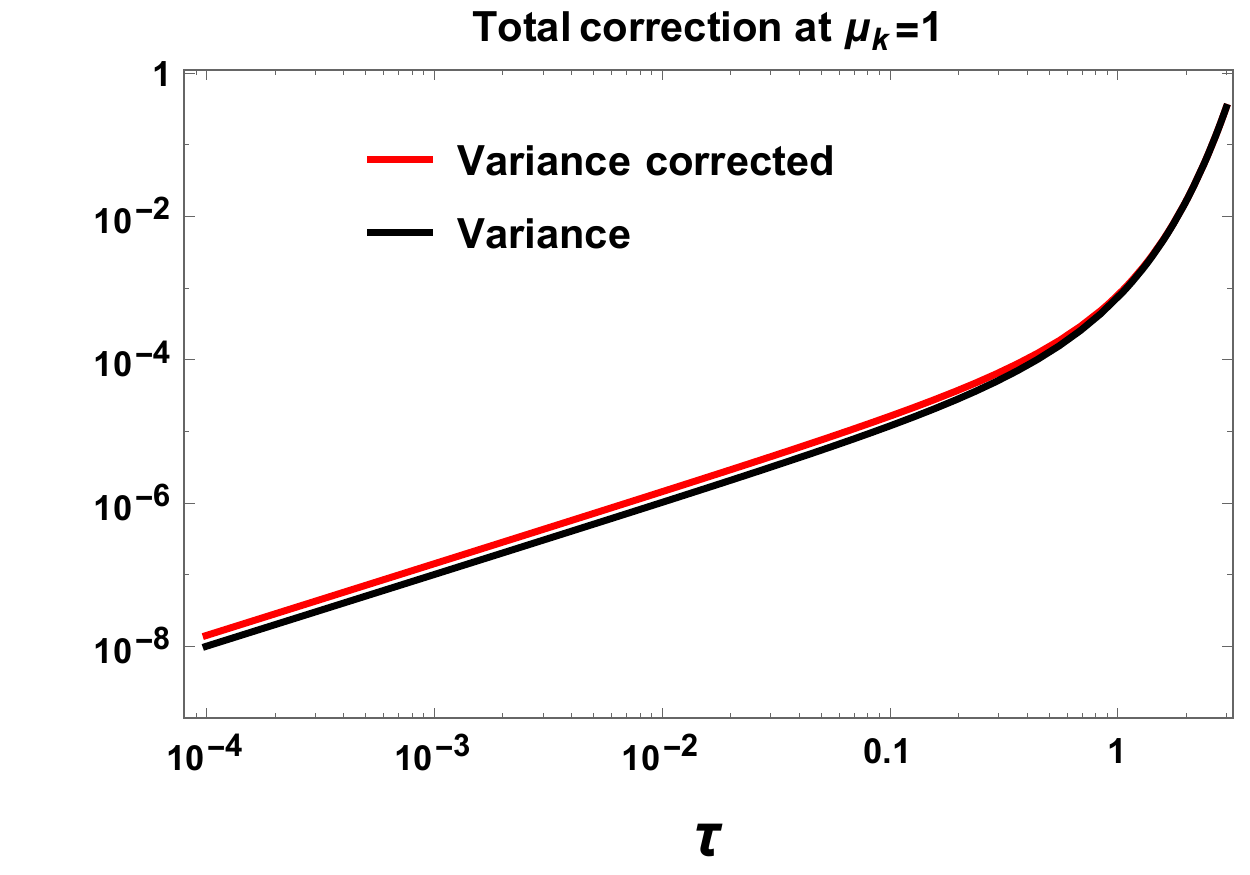}
    \label{sd1}
}
\subfigure[Time evolution of skewness.  ]{
    \includegraphics[width=8cm,height=5cm] {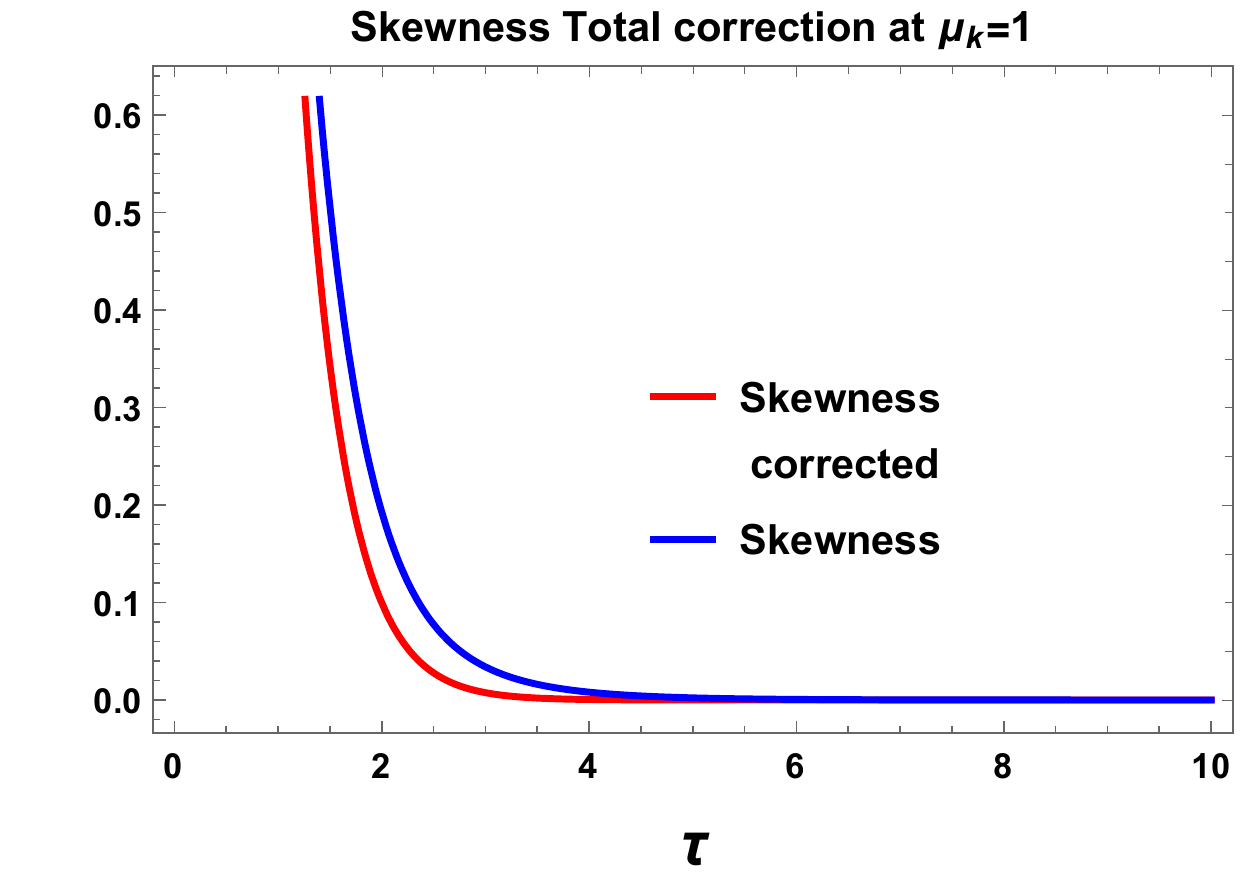}
    \label{SKRR1}
}
\subfigure[Time evolution of kurtosis. ]{
    \includegraphics[width=8cm,height=5cm] {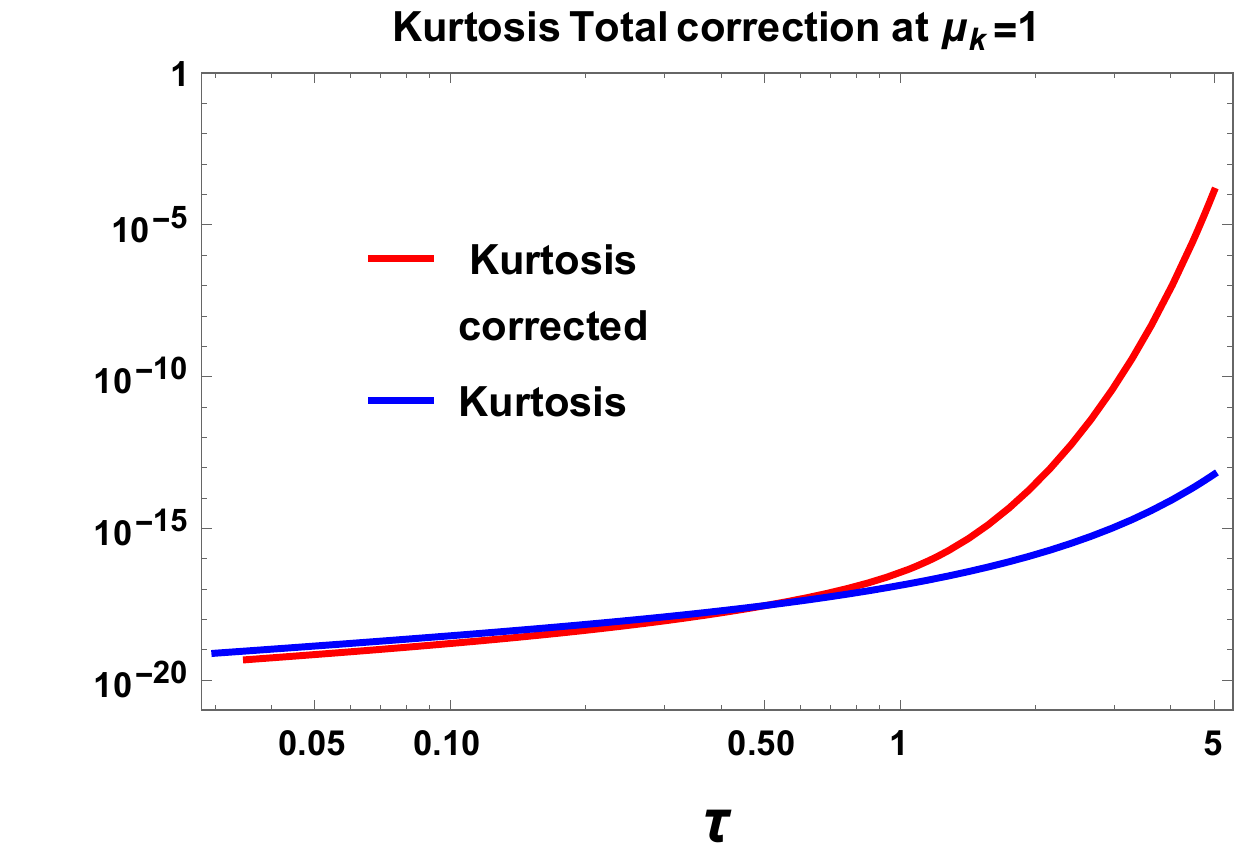}
    \label{STAT1x}
}
	\caption{Time evolution of variance, skewness and kurtosis computed from the probability distribution profile.  }
	\label{cvx}
\end{figure}

From Fig.~(\ref{Allcorrection}) the $P_{i}$ (i=1,2,3,4) denote the i-th order probability distribution. The order by order small corrections (fluctuations) from Gaussian profile support the quantum effects in stochastic particle production.
From the quantum corrected probability distribution we can further calculate different statistical moments using Eq~(\ref{ez1}). Calculating expression for $\langle n\rangle $,$\langle n^{2}\rangle$,$\langle n^{3}\rangle$ and $\langle n^{4}\rangle$ and standard deviation, skewness and kurtosis for a given time solidify the quantum nature as predicted earlier.

To compute the first moment of the occupation number we use the first order master evolution equation:
\bea \mu^{-1}_{k}  \partial_{\tau} \langle n\rangle = 1 + 2 \langle n\rangle~.\eea 
To compute the second moment we use first and second order master equations in two different orders:

\bea
	{\bf 1st~ order}:
	~~~
\mu^{-1}_{k}  \partial_{\tau} \langle n^2\rangle &=&4\langle n\rangle+6\langle n^2 \rangle,~~~~ \\
	{\bf 2nd~order}:~~~
\mu^{-2}_{k}\partial^{2}_{\tau} \langle n^2\rangle &=& 12 \langle n\rangle+12 \langle n^2\rangle+2.~~~
\eea
 Continuing in the same way one can similarly calculate third  and fourth moments corrected upto different orders. 
 
\begin{figure}[h]
\centering
\subfigure[Probability distribution for It$\hat{o}$.  ]{
    \includegraphics[width=8cm,height=7cm] {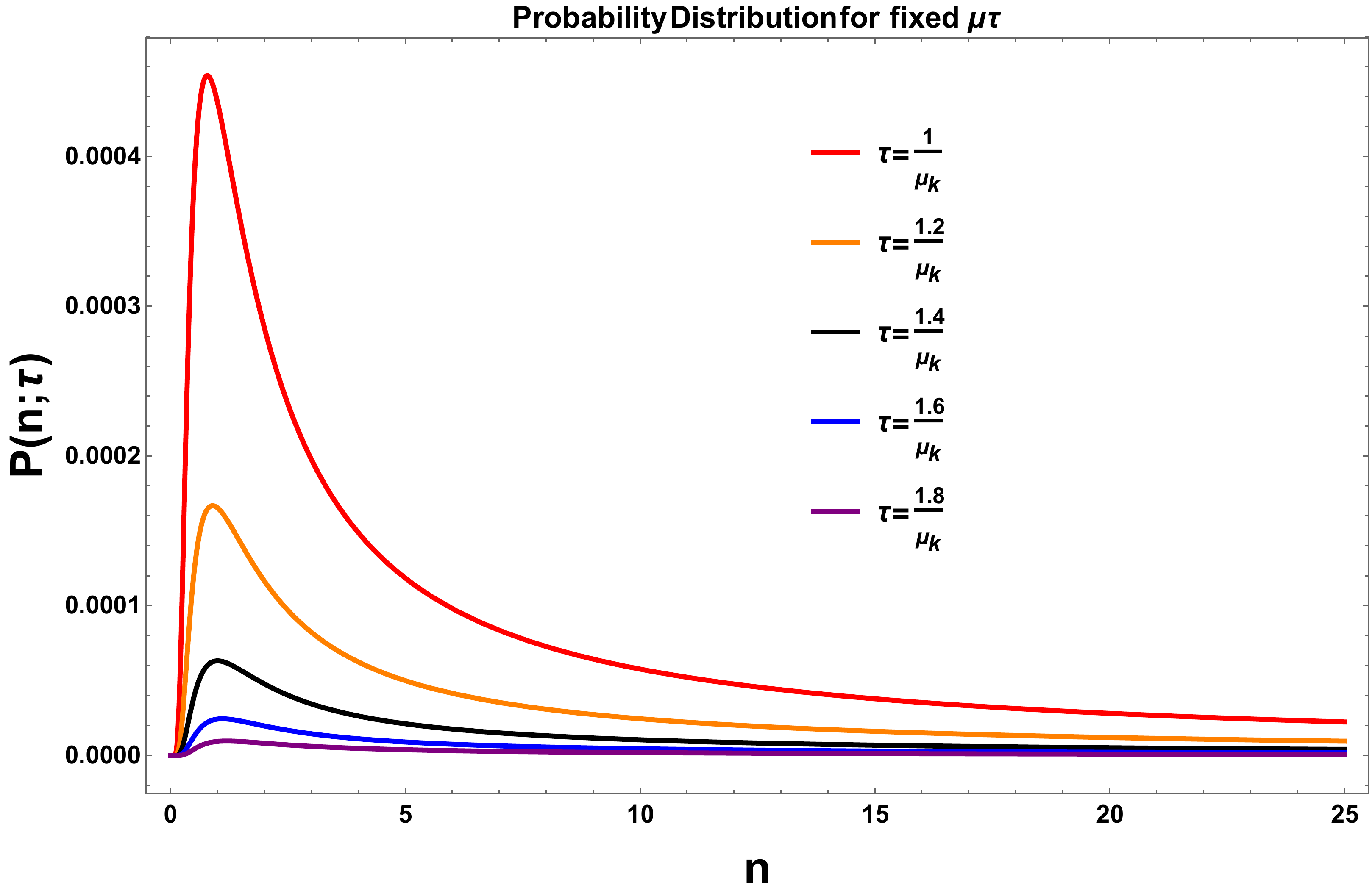}
    \label{ito1}
}
\subfigure[Probability distribution for Stratonovitch with $Q=1/2$.   ]{
    \includegraphics[width=8cm,height=7cm] {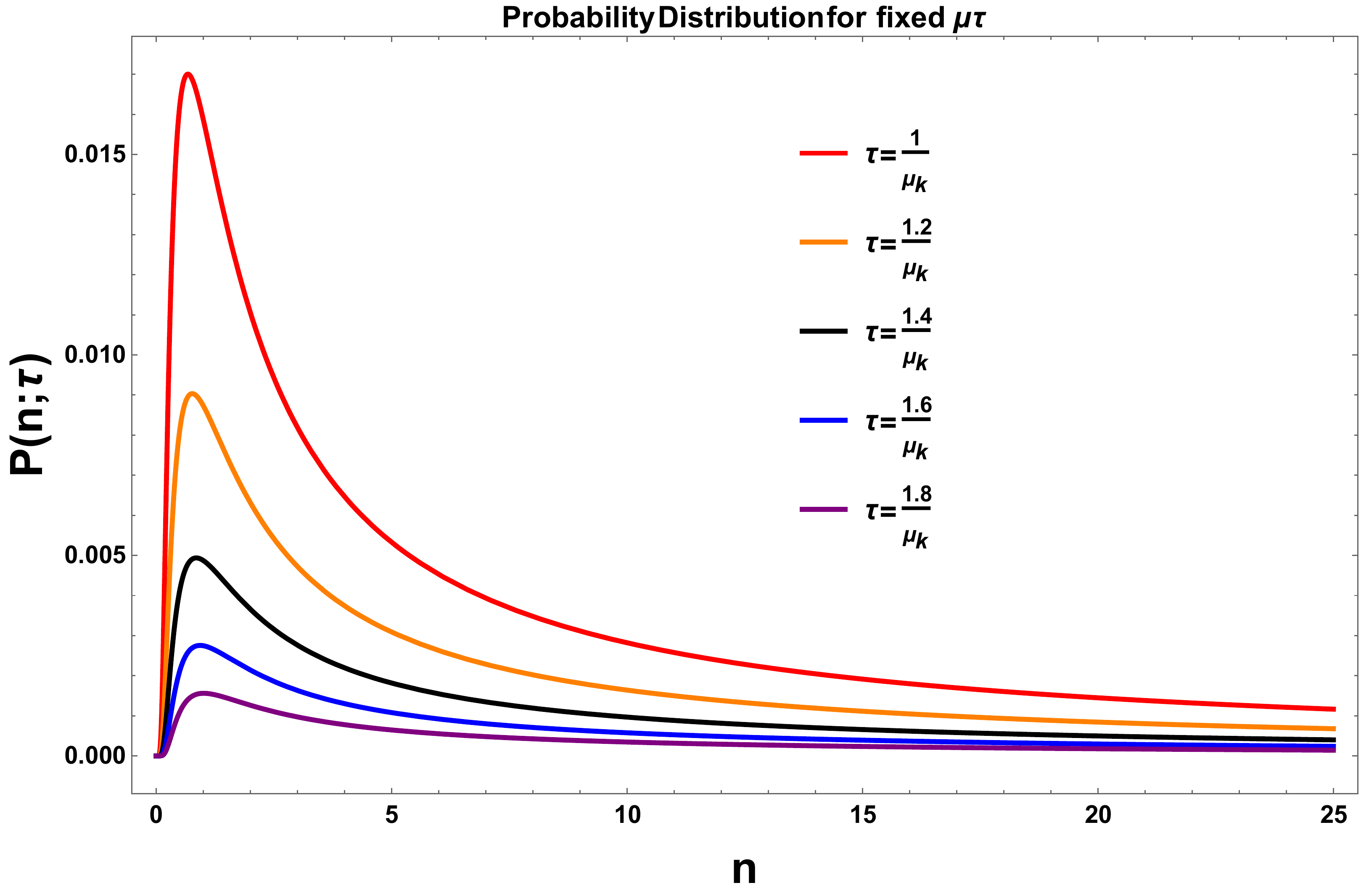}
    \label{STAT1}
}
	\caption{Variation of probability density function for It$\hat{o}$ and Stratonovitch with the occupation number per mode}
	\label{cvx}
\end{figure}

From Fig.~(\ref{sd1}) we obtain the large variance with increasing $\tau$. But the quantum corrected and uncorrected distribution have same variance at all time signifying that width of the peak is unchanged by the quantum effects.

Additionally, it is important to note that the computed probability distribution function has a long right tail a specific effect of positive skewness. Considering different order correction in $\langle n^{3} \rangle$ and standard deviation we calculate skewness with and without correction. Now from Fig.~(\ref{SKRR1}), we can say that the corrected  {\it Skewness} deviate significantly from the uncorrected one at low $\tau$ limit. But we can see that at higher time scale they overlap. So for particle production at initial time the skewness is dominant over uncorrected skewness. So the effects of quantum corrections are more clearly visible for initial time scale. Using the corrected $\langle n^{4}\rangle$ and standard deviation we calculate the kurtosis for particle production event which we have shown in Fig.~(\ref{STAT1x}). Here we have shown the quantum corrections are dominant at large time scale, but at low time scale both the corrected and uncorrected kurtosis overlap with each other.

\begin{figure*}[htb]
\centering
\subfigure[Probability distribution for $\beta=1/T=0.01$.  ]{
    \includegraphics[width=8cm,height=7cm] {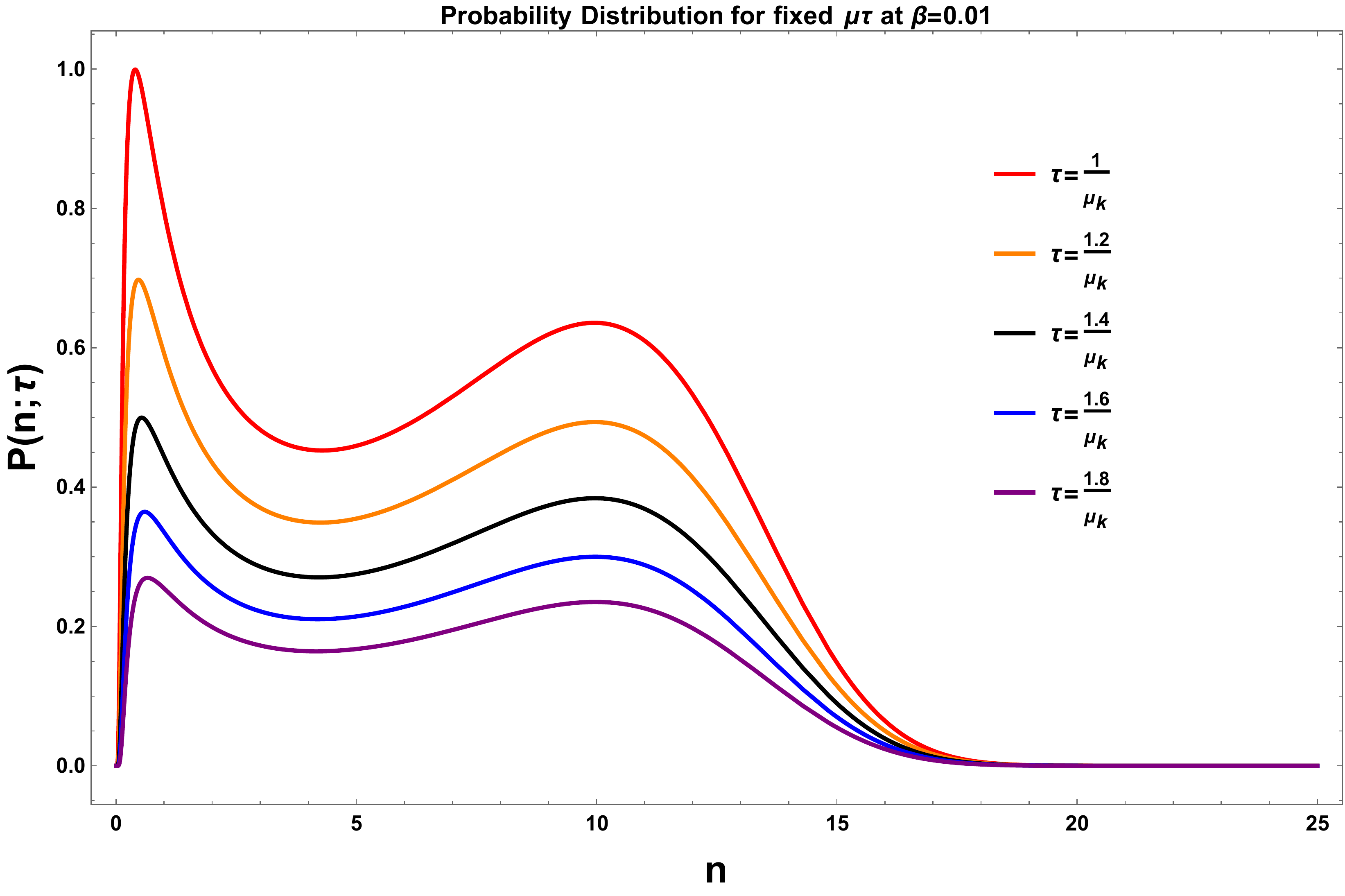}
    \label{4a}
}
\subfigure[Probability distribution for $\beta=1/T=100$.   ]{
    \includegraphics[width=8cm,height=7cm] {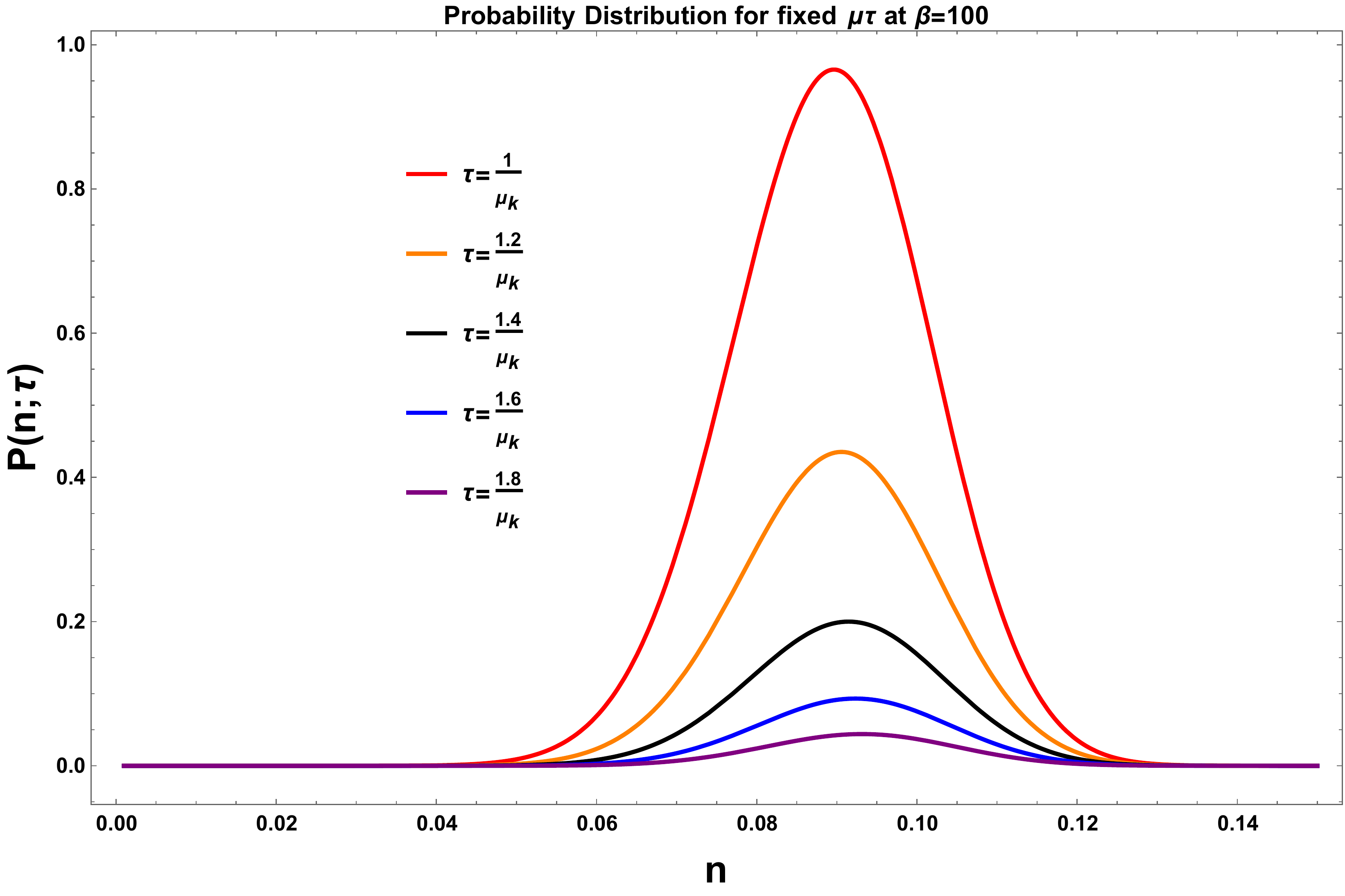}
    \label{4d}
}
	\caption{Variation of the probability density function with respect to the the occupation number per mode at different temperatures.}
	\label{r4e}
\end{figure*}
 From It$\hat{o}$ and Stratonovitch perspective the Fokker Planck equation can be expressed as:
 \bea
&&{\bf  It\hat{o}:}~~\partial_{\tau} P(n;\tau)=\partial^2_n(D(n)P(n;\tau)),\\
&&{\bf Stratonovitch:}\nonumber\\
&&\partial_{\tau} P(n;\tau)=\partial_n((D(n))^{1-Q}\partial_n((D(n))^Q P(n;\tau)))~~~~~~\eea
where $D(n)=n(n+1)$. Using this we get the following solution of probability distribution:
\bea &&{\bf  It\hat{o}:}~~ P(n,\tau)=\frac{\exp\left[-\frac{((4 n+2) \tau  \mu _k+n){}^2 }{4 n (n+1) \tau  \mu _k}\right]}{2 \sqrt{\pi } \sqrt{n (n+1) \tau  \mu _k}},\eea \bea
&&{\bf Stratonovitch:}~~\nonumber\\
&&  P(n,\tau)=\frac{1}{2 \sqrt{\mu_k \pi  \tau (n (n+1))^Q}}\nonumber\\
&&\times\exp \left[-\frac{\left(n^2 (n+1)+\mu_k \tau (2 n+1) Q (Q+1) (n (n+1))^Q\right)^2}{4 \mu_k \tau (n (n+1))^{Q+2}}\right].\nonumber\\
&&
\eea
The probability distribution function obtained from this have the same the log normal form.

From General perspective the Fokker Planck equation with effect of potential($U(n)$) can be expressed as:
\begin{widetext}
	\bea 
	&& \partial_n\left( D(n)~\partial_n \left(\exp\left(\frac{\beta V(n)}{2}\right)P(n;\tau)\right)\right)-U(n)\exp\left(\frac{\beta V(n)}{2}\right)P(n;\tau)=\partial_{\tau} \left(\exp\left(\frac{\beta V(n)}{2}\right)P(n;\tau)\right), \eea\end{widetext}
	where the effective potential at finite temperature can be expressed in terms of the diffusion function $D(n)$ and the specific model potential $V(n)$ for the number density of the created particles as:
	\begin{widetext}
	\bea
      U(n)&=&\left[\frac{\beta^2}{4}D(n)(\partial_n V(n))^2-\frac{\beta}{2}D(n)(\partial^2_n V(n))-\frac{\beta}{2}(\partial_n D(n))(\partial_n V(n))\right].
\eea
\end{widetext}

Choosing a specific form of the diffusion function, $D(n)=n(n+1)$ and the model potential for the number density of the created particles, $V(n)=n^{2}$ we get the following simplified expression for the probability distribution function at finite temperature:
\begin{widetext}\bea~&& P(n;\tau,\beta)=\frac{1}{2 \sqrt{\pi } \sqrt{n (n+1) \tau  \mu _k}}\times\exp\left[-\frac{(n-\mu _k (2 n \tau +\tau )){}^2}{4 n (n+1) \tau  \mu _k}-\frac{\beta  n^2}{2}-\beta  n \left\{n (\beta  n (n+1)-3)-2\right\}\right].~~~~
\eea\end{widetext}
This result is perfectly consistent as it can able to produce the previously obtained result in the limiting approximation, $\beta\rightarrow 0$ (or equivalently at $T\rightarrow \infty$). This happened because in this limit one can fix $U(n)\rightarrow 0$ and $\exp\left(\frac{\beta V(n)}{2}\right)\rightarrow 1$. As a result, we get,  \be P(n;\tau,\beta\rightarrow 0)=P(n;\tau),\ee where $P(n;\tau)$
is the probability distribution function which we have obtained in the {\bf It$\hat{o}$} prescription. 

From Fig.~(\ref{r4e}) of the probability distribution function we observe that for large value of occupation number  the distribution function decays to a finite saturation value. On the other hand for small occupation number we get peak in the distribution function for different values of $\mu_k\tau$. 
 
 With different order solutions of {\it Fokker Planck equation} we construct probability density function which explain the quantum nature in stochastic particle production scenario in early universe cosmology. Also the existance of higher order statistical moments of the probability density function. The present approach can extend to explain the semi-classical behaviour of particle production event and relating chaos to this approach eventually set a bound to the quantum randomness \cite{Choudhury:2018lcb}.
 
 From this non-gaussianity in stochastic particle production during inflation period we can connect it with the idea of non-gaussianity in finite universe. Considering interacting background field it will be possible to introduce the other non-linear and dissipative effects into the system introduced by the background itself and can be studied as open quantum system interacting with the defined background set-up \cite{Shandera:2017qkg}. Using more general statistical field theory along with using the well known {\bf It$\hat{o}$} and {\bf Stratnovitch} prescription in presence of general background potential at finite temperature the result for analytically obtained probability distribution function for particle creation can be further generalised for any system where randomness plays significant role within it.\\ \\
\\ \\
{\bf Acknowledgement:}~SC would like to thank Quantum Gravity and Unified Theory and Theoretical Cosmology
Group, Max Planck Institute for Gravitational Physics, Albert Einstein Institute for providing the Post-Doctoral Research Fellowship. SC take this opportunity to thank sincerely to
Jean-Luc Lehners for their constant support and inspiration. SC thank the organisers of Summer School on Cosmology 2018, ICTP, Trieste, 15 th Marcel Grossman Meeting, Rome, The European Einstein Toolkit meeting 2018, Centra, Instituto Superior Tecnico, Lisbon and The Universe as a Quantum Lab, APC, Paris for providing the local
hospitality during the work. We also thank all the members of our newly formed virtual group ``Quantum Structures of the Space-Time \& Matter" for elaborative discussions and suggestions to improve the presentation of the article. Last but not the least, we would like to acknowledge our debt to
the people belonging to the various part of the world for their generous and steady support for research in natural sciences.
\\

{\bf Note: } This project is the part of the non-profit virtual international research consortium ``Quantum Structures of the Space-Time \& Matter".
\\ 
{\bf Email address:}\\
$~~~~{}^{1}$sayantan.choudhury@aei.mpg.de,\\
$~~~~{}^{2}$am16ms058@iiserkol.ac.in



\end{document}